\documentclass[twocolumn]{article}
\usepackage{graphicx}

\begin{document}

\title{IR-stimulated visible fluorescence in pink and brown diamond}

\author{K S Byrne$^{1,\#}$, J G Chapman$^2$ and A N Luiten$^{1, 3}$\\
\quad\\
$^1$\small School of Physics, The University of Western Australia, Crawley 6009, Western Australia\\
$^2$\small Rio Tinto Diamonds, Perth 6000, Western Australia\\
$^3$\small Institute for Photonics and Advanced Sensing (IPAS) and \\\small The School of Chemistry and Physics, University of Adelaide, Adelaide 5005, South Australia}

\date{$^\#$\small Email: kbyrne@physics.uwa.edu.au}

\twocolumn[
\begin{@twocolumnfalse}
\maketitle
\begin{abstract}
 \normalsize Irradiation of natural pink and brown diamond by middle-ultraviolet light (photon energy $\epsilon \geq$4.1\,eV) is seen to induce anomalous fluorescence phenomena at N3 defect centres (structure N$_3$-V). When diamonds primed in this fashion are subsequently exposed to infrared light (even with a delay of many hours), a transient burst of blue N3 fluorescence is observed. The dependence of this IR-triggered fluorescence on pump wavelength and intensity suggest that this fluorescence phenomena is intrinsically related to pink diamond photochromism. An energy transfer process between N3 defects and other defect species can account for both the UV-induced fluorescence intensity changes, and the apparent optical upconversion of IR light. From this standpoint, we consider the implications of this N3 fluorescence behaviour for the current understanding of pink diamond photochromism kinetics.
\end{abstract}
\quad
\end{@twocolumnfalse}
]

\section{Introduction}
The large range of optically-active defect species found within diamond makes luminescence a frequently employed tool for determining which diamond defect species exist within a bulk sample~\cite{walker79, davies98, zaitsevbook}. Luminescence measurements can also provide insight into the dynamics of electronic processes involving diamond defects by, for instance, acting as a probe of a targeted defect centre's charge state during excitation of the centre~\cite{iakoubovskii00, liaugaudas09, beha12}. 

The N3 centre, with structure N$_3$-V and a zero-phonon line (ZPL) at 2.985\,eV (415\,nm), commonly occurs in natural diamonds, where it is readily identifiable from its luminescence spectrum~\cite{zaitsevbook}. In this paper, we discuss a regime of anomalous fluorescence behaviour in pink and brown Argyle diamonds, that has properties relating to both the N3 defect centre, and to the photochromism properties of Argyle pink diamond. To our knowledge these effects have not been described before. We present an experimental exploration of the anomalous fluorescence behaviours, in both the spectral and temporal domain. We propose that an energy transfer mechanism from the diamond defects responsible for pink diamond photochromism, to readily-fluorescing N3 centres, is responsible for the observed phenomena.

\section{Samples}
These experiments focused on a series of Argyle pink and brown diamond samples, across a range of colouration. Two samples displayed an obvious pink colouration, three samples were coloured predominantly brown with a pink undertone, and two samples were ``white''/colour-free. All samples were of type Ia - containing a concentration of nitrogen impurities in aggregated form, ranging between 50-1000\,ppm. 

The N3 defect is typically found throughout the bulk of Argyle pink diamonds, acting as a dominant feature in both absorption and fluorescence spectra~\cite{deljanin08, gaillou10}. All samples tested here displayed visible N3 absorption features, with one of the white diamonds displaying the strongest N3 concentration of all diamonds tested. These same samples have also been employed in recent photochromism experiments, to which we refer the reader for further details~\cite{byrne14a}. As we will show, there is a close connection between the anomalous fluorescence and photochromic behaviours.

\section{Results}
\subsection{UV- and IR-stimulated fluorescence}
\begin{figure*}
\centering
\includegraphics[width=0.9\textwidth]{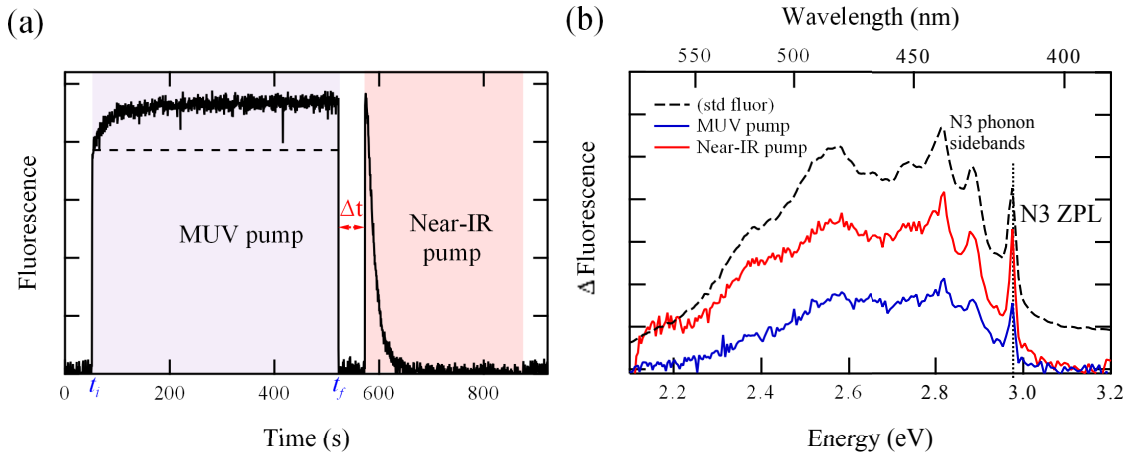}
   \caption{(a) Fluorescence intensity over time, under MUV and IR pumps. A constant-intensity MUV pump stimulates an increasing fluorescence intensity; after MUV-priming, IR light induces a transient, decaying burst of fluorescence. Contamination of the measurement by the IR pump light has been subtracted for clarity; the measurement is insensitive to the MUV pump light. (b) The measured change in MUV-induced fluorescence (from $t_i$ to $t_f$) mimics the initial fluorescence spectrum (scaled for clarity). The fluorescence excited by long-wavelength pumps also has the same spectrum.}
\label{N3fluordual}
\end{figure*}

In both pink and brown diamond samples, we observed that constant-intensity optical pumping in the middle ultraviolet (MUV, 4.1-6.2\,eV) leads to an emitted N3 fluorescence signal that increases in intensity over time, asymptotically approaching some higher level (see figure~\ref{N3fluordual}(a)). This is in strong contrast to pumping with longer-wavelength UV radiation, where the intensity of N3 fluorescence is constant. The threshold photon energy for this new behaviour to occur was found to be in the region of 4.1\,eV. The increased level of N3 fluorescence intensity is stable - re-exposure of the sample to the MUV pump after storing the sample in the dark for some time $\Delta t$ will immediately generate the same increased level of fluorescence. 

Once primed by MUV light in this fashion, the N3 fluorescence intensity could be promptly reset to its initial level by exposure to IR light. Intriguingly, this IR pump is also observed to trigger a transient emission of N3 fluorescence (which has a far higher photon energy than the triggering IR light). The intensity of this IR-stimulated fluorescence decays away exponentially with continued pumping, and IR light only generates such fluorescence when applied subsequent to MUV pumping of the diamond. 

Figure~\ref{N3fluordual}(a) shows the intensity of fluorescence emitted from a pink diamond over time, during both MUV-stimulated fluorescence priming and subsequent IR-stimulated transient decay, measured with a large-area silicon photodetector. The fluorescence signal was isolated using appropriate optical filtering prior to the detector. The signal-to-noise ratio of the measurement was improved by mechanically chopping the pump light to generate amplitude-modulated fluorescence. The photodiode signal was then  recovered using a lock-in amplifier.  

The spectrum of this fluorescence was recorded at various times within the pumping cycle using an UV-Visible spectrometer, integrating multiple measurement cycles to achieve a higher signal-to-noise ratio. The dashed line in figure~\ref{N3fluordual}(b) shows the standard fluorescence spectrum for these diamonds, showing both the characteristic ZPL (2.985\,eV) and phonon sidebands of the N3 defect. The blue trace on this figure shows the difference between the fluorescence spectrum at the start of MUV excitation $t_i$, to that at the end of the pumping period $t_f$ (when the system has reached the maximum fluorescence intensity). This difference spectrum is nearly identical to the spectral shape of the initial fluorescence, demonstrating that this increase in fluorescence intensity is merely an amplification of the usual fluorescence rather than the emergence of some other fluorescence signal. The spectrum of the IR-stimulated fluorescence (shown as the red trace on the figure) again mimics the N3 fluorescence profile.

While all the pink and brown samples showed this dynamic N3 fluorescence, the two white samples showed no such behaviour despite one of these containing the highest N3 concentration of all tested samples. Thus these phenomena are not intrinsic to N3, but depend also upon the presence of some other defect that acts in concert with N3 to produce the observed effects.

\subsection{IR-stimulated fluorescence yield and decay rates}
\begin{figure}
\centering
\includegraphics[width=0.45\textwidth]{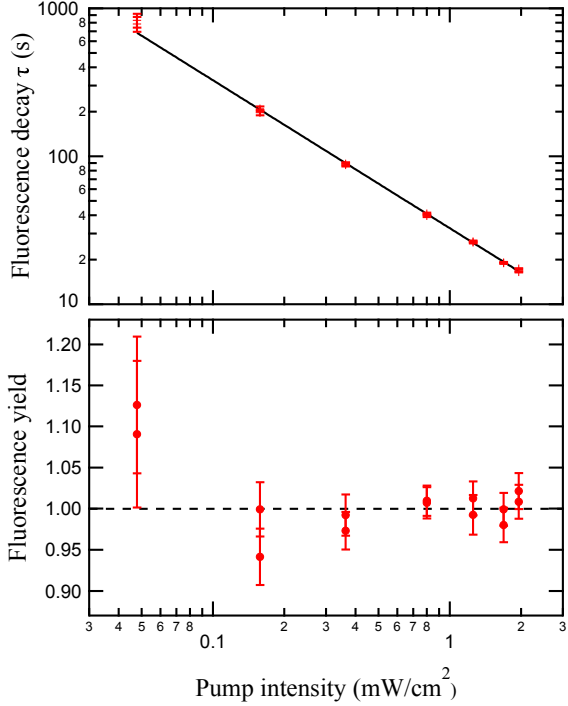}
   \caption{The decay time-constant for IR-stimulated fluorescence is inversely proportional to IR pump intensity, but the total yield of fluorescence (normalised scale) is constant.}
\label{pumpintensity}
\end{figure}

The total integrated fluorescent output (yield) generated by IR-pumping is constant regardless of the IR pump intensity, or whether the pump is interrupted and resumed. Figure~\ref{pumpintensity} shows measurements of IR-induced fluorescence for a MUV-primed diamond subsequently pumped with a 2.0\,eV LED at various intensities. The decay rate of the IR-stimulated fluorescence is inversely proportional to the intensity of the IR pump, however the fluorescence yield remains unchanged. Pumps at other wavelengths show equivalent results. 

\begin{figure}
\centering
\includegraphics[width=0.45\textwidth]{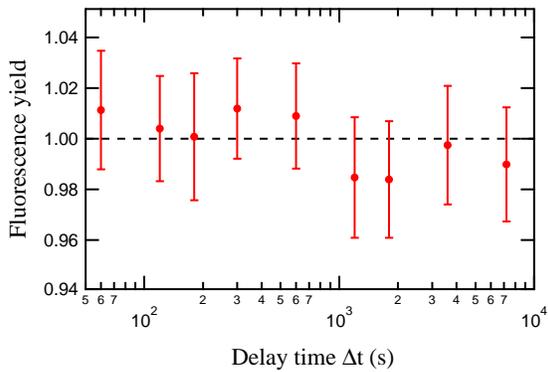}
   \caption{The yield of the IR-stimulated fluorescence process (normalised scale) is independent of delay time $\Delta t$ on timescales up to hours.}
\label{deltat}
\end{figure}

The ``storage time'' of the system was also investigated, by priming a diamond with MUV light and then waiting some time $\Delta t$ before applying an IR pump of fixed intensity. As evident in figure~\ref{deltat}, no net loss in the output fluorescence yield of the system was detected, even when storing a UV-primed diamond in the dark for several hours before IR exposure. It appears that at room temperature and in the dark, the UV-primed system is entirely metastable.

\subsection{Excitation threshold for IR-stimulated fluorescence}
\label{cssection}
\begin{figure}
\centering
\includegraphics[width=0.45\textwidth]{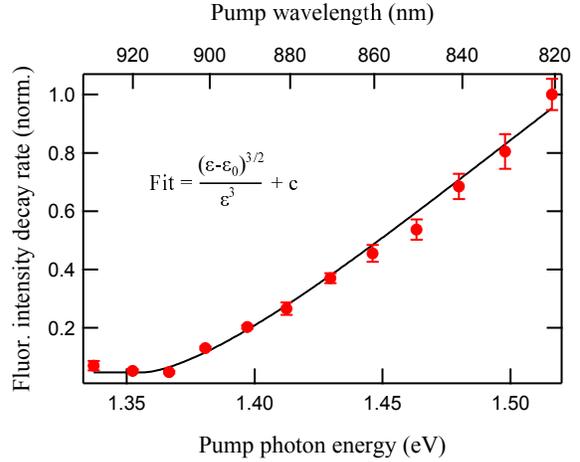}
   \caption{Rate of fluorescence decay normalised to incident pump photon count. Fitting the Lucovsky cross-section to the data shows a clear photon energy threshold at $\epsilon_0=$1.357\,eV.}
\label{fluorwl}
\end{figure}

To determine the photon energy threshold for the emission of IR-stimulated fluorescence, the rate of fluorescence decay was measured by using a Ti:Sapphire laser as a tunable IR pump. The resulting fluorescence decay rate was normalised to the incident pump photon flux to reflect the cross-section of the IR-stimulated interaction (figure~\ref{fluorwl}). A photon energy threshold is evident, which was determined to lie at $\epsilon_0=$1.357$\pm$0.001\,eV by fitting the Lucovsky cross-section~\cite{lucovsky65} to the data, together with a small constant term that we attribute to weak fluorescence excitation induced by background light leaking onto the sample.

\section{Discussion}
Photochromism of the N3 centre via optically-induced charge transfer has been documented in literature~\cite{mita97}, albeit resticted to treated type-Ib diamonds. However, charge trapping and photoionisation at N3 defects is unlikely to be responsible for the phenomena described above, given that the white diamonds did not show this dynamic fluorescence. Additionally, although the photon energy necessary to excite N3 fluorescence (2.985\,eV) is well above that required to photoionise the pink colour centre and thus induce photochromic bleaching of a pink diamond (2.7\,eV~\cite{byrne12b}), no variation is seen in either the spectrum or the quantity of N3 fluorescence when near-ultraviolet pumps (e.g. 3-3.5\,eV) are applied to our diamonds, indicating that the N3 centres are not trapping electrons photoionised from the pink colour centres.

\begin{figure*}
\centering
\includegraphics[width=0.6\textwidth]{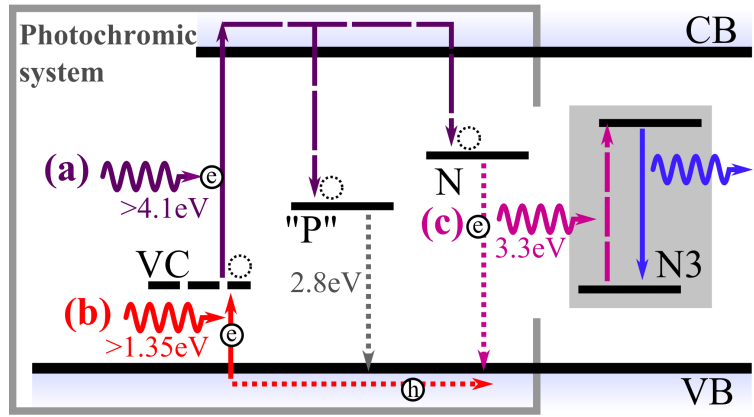}
   \caption{N3 fluorescence may be generated via internal energy transfer from other centres participating in the IR-induced photochromism transitions: (a) MUV excitation ionises vacancy cluster electrons, which are trapped at other centres participating in the photochromic process. (b) IR light excites valence electrons into empty vacancy cluster states. (c) Electrons trapped at other defects recombine with the created valence holes, releasing sufficient energy to excite a nearby fluorescent N3 centre.}
\label{levels3}
\end{figure*}

The photochromic response of pink and brown diamonds to UV and IR pumps~\cite{byrne14a} displays strong similarity to the fluorescence behaviours described here. A new regime of pink/brown photochromism, believed to correspond to the ionisation of electrons at vacancy cluster defects, appears under the influence of MUV sources with photon energies $>$4.1\,eV; electrons ionised from the vacancy clusters may then be trapped at other defects such as substitutional nitrogen and pink colour centres (see figure~\ref{levels3}). Later exposure of the diamond to IR light results in further photochromic change consistent with the excitation of valence electrons to empty vacancy cluster states, and subsequent recombination of electrons trapped at other centres into the new valence holes. 

The cross-section of the IR-stimulated photochromic change (figure 5 in~\cite{byrne14a}), with a photon-energy threshold measured to lie at 1.353$\pm$0.001\,eV, is essentially equivalent to the excitation cross-section measured here for IR-stimulated fluorescence (figure~\ref{fluorwl}, threshold at 1.357\,eV). The very close similarity between the measured MUV and IR threshold energies in photochromism and fluorescence experiments - as well as the observation that IR sensitivity in both fluorescence and photochromism is a secondary phenomenon, dependent on earlier exposure of the sample to MUV light - suggests that the same phenomena underlie both sets of observations.  

Recombination of electrons trapped at pink colour centres and substitutional nitrogen centres would liberate energies of 2.8\,eV and 3.3\,eV respectively. The latter is easily of sufficient energy to excite an N3 centre from its ground state, and the N3 centre has been seen to participate in energy transfer interactions with other defect species in natural diamond samples~\cite{thomaz78}. We thus suggest that these recombination transitions, at the centres participating in the photochromic electron exchange, result in energy transfer to N3 centres, whether radiatively or via some local energy exchange process, resulting in the transient blue-light fluorescence signal which is emitted under IR excitation. From this model, we can understand the total fluorescence intensity generated by MUV pumps, measured in figure~\ref{N3fluordual}(a), as being due to two separate processes: MUV light directly exciting N3 centres and generating a steady output (indicated by the dotted line) and, simultaneously, recombination events at other defects in the crystal generating additional N3 fluorescence via energy exchange. 

\begin{figure}
\centering
\includegraphics[width=0.45\textwidth]{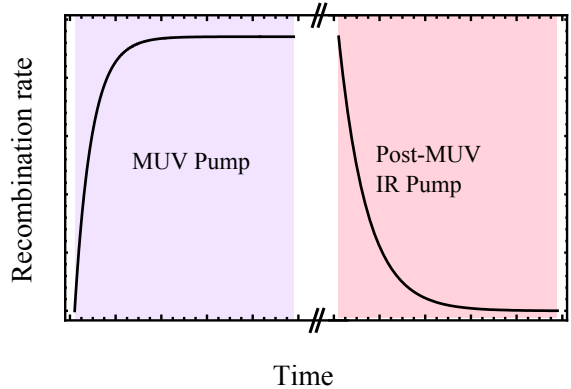}
   \caption{The rate of recombination events from photochromism traps mimics the dynamics of the fluorescence response seen at the N3 centre.}
\label{chargedecay}
\end{figure}

Using a rate equation model of the energy level system shown in figure~\ref{levels3}, we calculate the expected rate at which electrons decay from the N and P centres over time as a result of optical stimulation with MUV and IR pumps. The result of this modelling is shown in figure~\ref{chargedecay}, using the count of recombinations as a proxy for the temporal variation in N3 fluorescence. The model only assumes that a constant fraction of the recombination events will excite the fluorescence process. The calculation predicts that under MUV excitation, the rate of recombination events rises approximately exponentially, asymptotically approaching some constant value in agreement with experimental observations. When the MUV pump is switched off, all traps are thermally stable, and no electronic transitions occur. Upon IR irradiation, there is an immediate burst of recombination events as holes are created in the valence band; the number of recombination events per unit time decays exponentially to zero as the population of valence holes is exhausted, once again in close agreement with observations. Comparing the calculations shown in figure~\ref{chargedecay} with the experimental observations in figure~\ref{N3fluordual}(a) demonstrates this agreement.

\section{Conclusion}
We have observed anomalous fluorescence from N3 defect centres in pink/brown Argyle diamonds. The photon energy threshold and time-dependence of these processes imply a strong link with the photochromism system that is also seen in pink/brown diamond. We deem it likely that the defects that drive the photochromic response in the diamond (and hence also give the diamond its pink colouration) have an energy coupling to nearby N3 centres which result in these new fluorescence behaviours.

\section{Acknowledgments}
The authors would like to thank Dr. Nigel Spooner, Dr. James Anstie and Dr. Phil Light from the University of Adelaide, for both technical assistance and helpful discussions. This research was supported by the South Australian Government through the Premier's Science and Research Fund as well as by the Australian Research Council through FT0991631.

\bibliographystyle{unsrt}
\bibliography{fluorbib}

\end{document}